\documentclass{cupconf}
\usepackage{epsfig}
\usepackage{rotating}

\let\realverbatim=\verbatim
\let\realendverbatim=\endverbatim
\renewcommand\verbatim{\par\addvspace{6pt plus 2pt minus 1pt}\realverbatim}
\renewcommand\endverbatim{\realendverbatim\addvspace{6pt plus 2pt minus 1pt}}
\makeatletter
\newcommand\verbsize{\@setfontsize\verbsize{10}\@xiipt}
\renewcommand\verbatim@font{\verbsize\normalfont\ttfamily}
\makeatother
  \checkfont{eurm10}
  \iffontfound
    \IfFileExists{upmath.sty}
      {\typeout{^^JFound AMS Euler Roman fonts on the system,
                   using the 'upmath' package.^^J}%
       \usepackage{upmath}}
      {\typeout{^^JFound AMS Euler Roman fonts on the system, but you
                   don't seem to have the}%
       \typeout{'upmath' package installed. CUPconf.cls can take advantage
                 of these fonts,^^Jif you use 'upmath' package.^^J}%
      }
  \else
  \fi

  \checkfont{msam10}
  \iffontfound
    \IfFileExists{amssymb.sty}
      {\typeout{^^JFound AMS Symbol fonts on the system, using the
                'amssymb' package.^^J}%
       \usepackage{amssymb}%

      }{}
  \fi

  \IfFileExists{amsbsy.sty}
    {\typeout{^^JFound the 'amsbsy' package on the system, using it.^^J}%
     \usepackage{amsbsy}}
    {}

\newsavebox{\astrutbox}
\sbox{\astrutbox}{\rule[-5pt]{0pt}{20pt}}

  \title[Cosmological Parameters From Radio Galaxies]{Cosmological Parameters
and Quintessence From Radio Galaxies}

  \author[Daly \& Guerra]{%
  R\ls U\ls T\ls H\ns 
  A.\ns 
  D\ls A\ls L\ls Y$^{1,3}$\ns 
  \and\ns
  E\ls R\ls I\ls C\ls K\ns
  J.\ns 
  G\ls U\ls E\ls R\ls R\ls A$^2$}

  \affiliation{$^1$Department of Physics,  Berks-Lehigh Valley
College, Penn State University,  P. O. Box 7009, Reading,
PA 19610-6009, USA; rdaly@psu.edu
\\[\affilskip]
    $^2$Department of Chemistry \& Physics, Rowan University, 
    Glassboro, NJ 08028-1701, USA; guerra@scherzo.rowan.edu
\\[\affilskip]
    $^3$NSF National Young Investigator}

\begin{document}

\maketitle

\begin{abstract}

FRIIb radio galaxies provide a tool to determine
the coordinate distance to sources at redshifts from zero to 
two.  The coordinate distance depends on the present values
of global cosmological 
parameters, quintessence, and the equation of state of 
quintessence.  The coordinate distance 
provides one of the cleanest determinations
of global cosmological parameters because it
does not 
depend on the clustering properties of any
of the mass-energy components present in the universe.

Two complementary methods that provide direct determinations
of the coordinate distance to sources with redshifts out to one 
or two are the modified standard yardstick method utilizing
FRIIb radio galaxies, and the modified standard candle
method utilizing type Ia  supernovae.  These two methods
are compared here, and are found to be complementary in 
many ways.  The two methods do differ in some regards; perhaps
the most significant difference is that the radio galaxy
method is completely independent of the local distance scale
and independent of the properties of local sources, 
while the supernovae method is very closely tied to the local
distance scale and the properties of local sources.  

FRIIb radio galaxies provide one 
of the very few reliable probes of the coordinate distance to
sources with redshifts out to two.   This
method indicates that the current value of the density parameter in 
non-relativistic matter, $\Omega_m$, must be low, irrespective
of whether the universe is spatially flat, and of whether a 
significant cosmological constant or quintessence pervades the 
universe at the present epoch.  

The effect of quintessence, with equation
of state $w$, is considered.  
FRIIb radio galaxies indicate that the universe is
currently accelerating in its expansion 
if the primary components of the universe
at the present epoch are non-relativistic
matter and quintessence, and the 
universe is spatially flat.

\end{abstract}

\firstsection % if your document starts with a section,
\section{Introduction}

Current values of global cosmological parameters that 
control and describe
the state and expansion rate of the universe are still not known with 
certainty.  The components of the universe at the present epoch
can be put into three categories:  
non-relativistic matter; photons and neutrinos
(important in the early universe); and a third component that
has yet to be identified, and may be a 
cosmological constant, quintessence, space curvature, or something else.  

Non-relativistic matter includes baryons and the dark matter known
to cluster with galaxies and galaxy clusters; the total, normalized,
mass density contributed by this component at present is
$\Omega_m$.  Non-relativistic matter is known to 
play an important role in the 
expansion rate of the universe at the present epoch.  Photons 
that make up the cosmic microwave background and neutrinos
produced in the early universe are known with some
certainty, but do not contribute significantly to the 
mass-energy density at present, and hence do not play an
important role in controlling the expansion rate of the
universe at the present epoch.  

There are numerous indications that $\Omega_m$ is low;
in fact, radio galaxies alone indicate that $\Omega_m$
must be less than about 0.63 at about 
95 \% confidence, and radio galaxies alone
indicate that $\Omega_m$ equal to unity is ruled
out at about 99 \% confidence (Daly \& Guerra 2001a [DG01a];
Guerra, Daly, \& Wan 2000 [GDW00]).  This includes
possible contributes from space curvature, a cosmological
constant, or quintessence.  Many methods indicate that
$\Omega_m$ is low (e.g. Turner \& White 1997; Perlmutter, 
Turner, \& White 1999; Wang et al. 2000).

If $\Omega_m < 1$, then there must be a third component,
and this component must play a significant role in 
determining the expansion rate of the universe at the present epoch.
This component could be space curvature, or could be
a component of mass-energy with an equation of state
$w = P/\rho$ that is different from non-relativistic
matter, which has $w=0$; here $P$ is the pressure
of the component, and $\rho$ its mass-energy density. 
Hopefully, there is only one unknown component that is 
significant at the present epoch.  

Recent observations of fluctuations of the microwave background
radiation at the last scattering surface indicate that
space curvature is close to zero (de Bernardis et al.\ 2000, Balbi et al.\
2000, Bond et al.\ 2000).  
This simplifies the determination of the third
component. A general form for the third component,
referred to as quintessence, 
allows constraints on both the normalized mass-energy
density $\Omega_Q = 1- \Omega_m$, and equation of state $w$.
Constraints imposed by FRIIb radio galaxies on quintessence are presented
in section 4. FRIIb radio galaxies are the most powerful FRII 
sources; they have regular radio bridge structure indicating
an average growth rate that is supersonic (see Daly 2001; GDW00, and
Wan, Daly, \& Guerra 2000 [WDG00]).  

The outline of the paper is as follows.  In section 2, the
radio galaxy and supernova methods are compared.  In section 3,
the radio galaxy method is described more fully.  The method
is applied to a spatially flat universe with quintessence in
section 4; it is shown that the radio galaxy method indicates 
that the expansion of the universe is accelerating at present.
In addition, the radio galaxy method is applied to 
a universe that may have space curvature,
a cosmological constant, and non-relativistic matter; it is shown
that radio galaxies indicate that $\Omega_m$ must be low.  

The application of radio galaxies as a modified standard
yardstick not only yields constraints on global cosmological
parameters and quintessence, but also yields constraints on
models of energy extraction from massive black holes, as  
discussed in section 5.  FRIIb radio sources also allow a 
determination of the pressure, density, and temperature of the gas 
around the source, as discussed in section 6.  
Studies of FRIIb sources indicate that they
are in the cores of clusters or proto-clusters of galaxies, and
thus provide a way to study evolution of structure, and a separate
way to constrain cosmological parameters.  Conclusions are
presented in section 7.

\begin{table}
\caption{Comparison of Supernova and Radio Galaxy Methods}
\label{Table1}
\begin{center}
\begin{minipage}{10.0 cm}

\begin{tabular}{@{}cccl@{}}
{}&{}\\
{\bf Supernovae} & {\bf Radio Galaxies} \\
        {}&{}\\
{Type SNIa} & {Type FRIIb} \\
$\propto (a_or)^{2.0}$ & $\propto (a_or)^{1.6}$\\
$0<z <1$ & $0<z<2$ \\
$\sim 100$ sources&20 sources (70 in parent pop.)\\
modified standard candle & modified standard yardstick\\
light curve $\Longrightarrow$ peak luminosity&radio bridge $\Longrightarrow$
average length\\
empirical relation  & 
physical relation \\
normalized at $z=0$ & not normalized at $z=0$\\
(depends on local distance scale) & (independent of local distance scale)\\
depends on local source properties & independent of local sources\\
{} & $\Omega_m$ is low \\
universe is accelerating & universe is accelerating ($\sim$ 2 $\sigma$)\\
some theoretical understanding  & 
good theoretical understanding \\
well-tested empirically & needs more empirical testing\\
   \end{tabular}
     \end{minipage}
    \end{center}
    \end{table}

\section{Comparison of Radio Galaxy and Supernova Methods}

Constraints on global cosmological parameters through the
determination of the coordinate distance to high-redshift sources
is particularly clean since it depends only on global cosmological
parameters such as $\Omega_m$ and $\Omega_{\Lambda}$ allowing
for space curvature, or $\Omega_m$ and the equation of state
of quintessence $w$ assuming zero space curvature and allowing
for quintessence.  The coordinate distance does {\it not} depend
on how the mass is clustered, the properties of the 
initial fluctuations, 
how density fluctuations evolve, whether baryonic
and clustered dark matter are biased, and a whole host
of issues that confound other methods of determining 
global cosmological parameters.  The only assumption is
that the different mass-energy components are
homogeneous and isotropic on large scales, 
scales much smaller than the scale of the coordinate
distance being determined.  

Two methods currently being used to determine the coordinate distance
to high-redshift sources 
are the radio galaxy method and the supernova method.  
Table 1 lists a comparison of key aspects of each method.
The supernova method uses type Ia supernovae as a modified standard
candle, as summarized in the papers of Riess et al.\ (1998)
and Perlmutter et al.\ (1999).
The rate of decline of the light
curve is used to predict the peak luminosity of
a supernova; the predicted peak luminosity and the 
observed peak flux density are used to determine
$(a_or)^2$, where $(a_or)$ is the coordinate distance
to the source.  (Recall that the luminosity distance
is $d_L = (a_or)(1+z)$, and the source redshifts are all
known.)

\begin{figure}
\centering
\epsfig{file=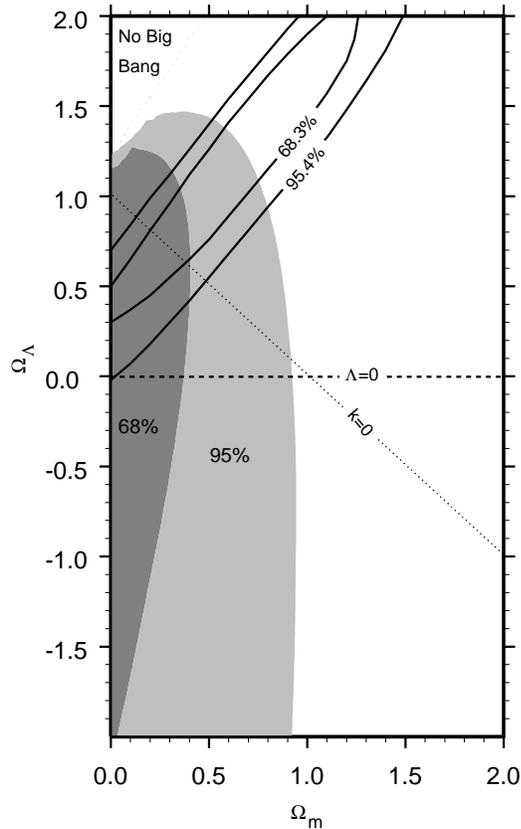,height=5truein}
  \caption{Two-dimensional constraints obtained using radio galaxies 
(shaded regions) compared with those from obtained by the 
supernovae cosmology team 
using supernovae (solid lines)(see Perlmutter et al. 1999 and
GDW00).  One-dimensional
constraints obtained using radio galaxies are discussed in the text.}
  \label{sncoscomp}
\end{figure}

\begin{figure}
\centering
\epsfig{file=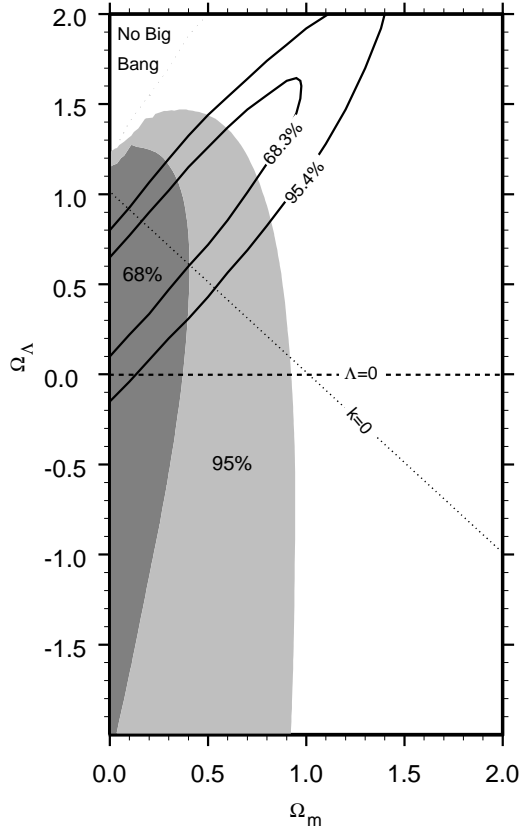,height=5truein}
  \caption{Two-dimensional constraints obtained using
radio galaxies (shaded regions) 
compared with those obtained by the high-redshift
supernovae team using supernovae (solid lines)
(see Riess et al. 1998 and GDW00).  One-dimensional
constraints obtained using radio galaxies are discussed in the text.}
  \label{riesscomp}
\end{figure}

\begin{figure}
\centering
\epsfig{file=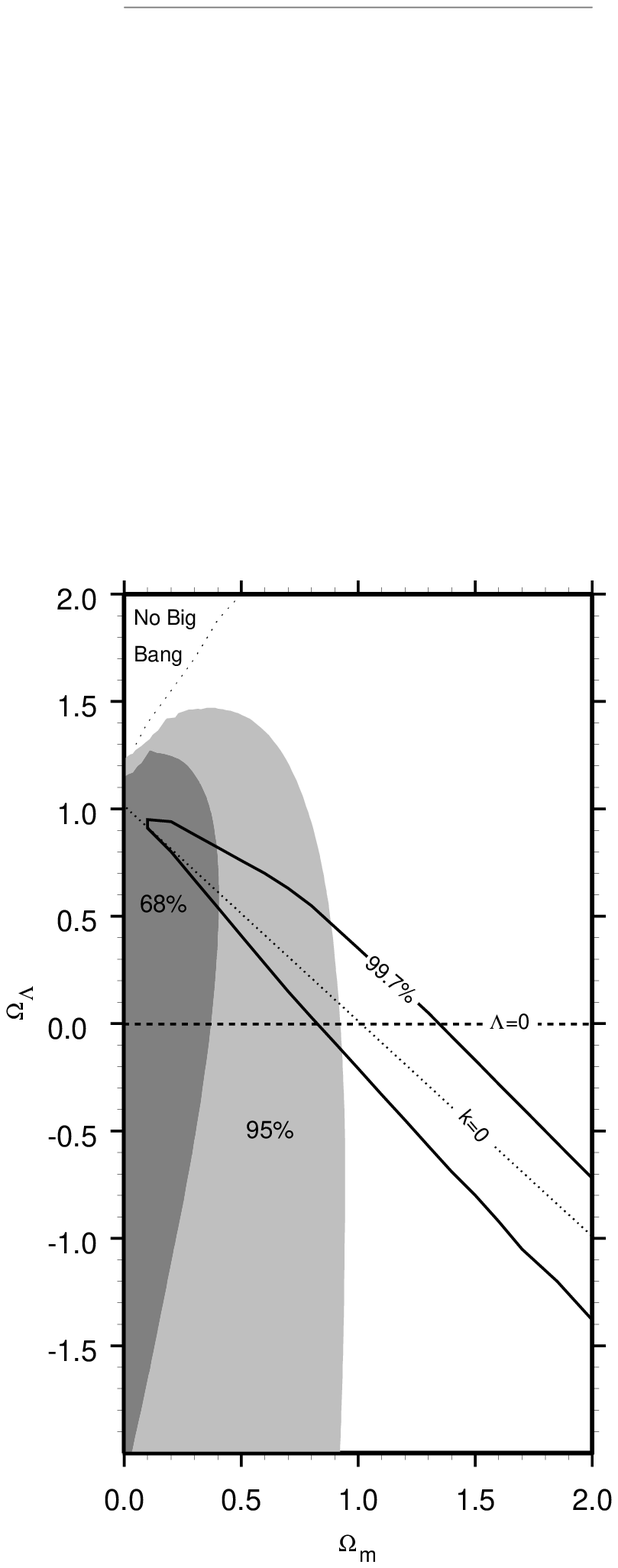,height=5truein}
\caption{Two-dimensional constraints from radio galaxies 
(shaded regions) compared with those 
from CMB anisotropy (solid lines) (see Bond et al. 
2000 and GDW00).  The CMB anisotropy measurements
indicate that k = 0.}
\label{cmb}
\end{figure}

The radio galaxy method uses FRIIb radio galaxies as a
modified standard yardstick, and is described in more
detail in section 3.  
The method relies on a comparison
between an individual source and the properties of the parent
population at the source redshift.  The radio surface brightness
and width of the radio source can be used to predict the 
maximum source size, measured using the largest angular
size of the source or the separation between the 
radio hotspots.  The average source size $D_*$ 
is just half of the maximum source size predicted 
using the model.  This
determination of the average source length does not
depend on the observed length of the source, $D$, and 
depends on the coordinate distance to the $(-2\beta/3)+(4/7)$ power:
$D_* \propto (a_or)^{(-2\beta/3)+(4/7)}$. The predicted average
size of a given source is equated to the average size of
all FRIIb radio galaxies at similar redshift $<D>$, and,
of course, $<D> \propto (a_or)$.  Clearly the two measures of
the average source size must be equal, or their ratio must
be a constant, independent of redshift:
$<D>/D_* = $ constant.  And, their ratio
depends on  $(a_or)^{(2\beta/3)+(3/7)}$.  Thus, this method allows
a determination of the model parameter $\beta$ and global
cosmological parameters.  It turns out that $\beta$ is relevant
to models of jet formation and energy extraction from AGN,
as discussed in section 5, and by Daly \& Guerra (2001b) [DG01b].  

The radio galaxy and supernova methods are complementary, as shown
in Table 1.  They have a similar dependence on the coordinate
distance; the power listed for radio galaxies is for a 
value of $\beta$ of 1.75, but this power does not change
by very much given the range of $\beta$ allowed by the 
data ($\beta = 1.75 \pm 0.25$).  They cover similar
redshift ranges, though the radio galaxy method does go
to redshifts of two.  They have a similar number of sources,
though the supernova method does have more sources than
the radio galaxy method, and thus is much more well tested
empirically, as noted in the
table.  The supernova method relies on a relation that
is derived empirically, while
the radio galaxy method relies on a relation that is derived 
using physical and theoretical arguments (see section 3).  The supernova
method is normalized at zero redshift, and depends quite
strongly on the local distance scale and properties of
local supernova.  The radio galaxy method is not
normalized at zero redshift, is completely independent 
of the local distance scale, and is independent of the
properties of local sources.  GDW00 show
that the analysis can be carried using sources with   
redshifts greater than 0.3 only, and the results are identical
to those obtained including sources with redshifts less
than 0.3; this is also shown here in figures
4, 5, and 8.  The radio galaxy method indicates that $\Omega_m$
must be low irrespective of the properties of the unknown 
component at zero redshift.  This method also implies
that if space curvature is zero, then it is quite likely
(84 \% confidence) that the universe is accelerating at
the present epoch (see figure 9).  The supernova method indicates that
the universe is accelerating at the present epoch.  The supernova
method is well-tested empirically, but needs more theoretical
understanding, while the radio galaxy method is better
understood theoretically, and needs more empirical testing.

The supernova method relies on relatively short-lived optical
events on the scale of stars, while the radio galaxy method
relies on much more long-lived radio events on scales larger
than the scale of galaxies.  Thus, any possible selection 
effects or unknown systematic errors must
be completely different for the two methods.  The fact that they
yield such similar results suggests that both are accurate,
and are not plagued by unidentified errors.  

The two-dimensional results obtained
using radio galaxies are compared with those obtained using 
supernova and the cosmic microwave background 
allowing for space curvature, a cosmological constant,
and non-relativistic matter, and are shown in figures 1, 2,
and 3.  Note, that the contours shown are a joint probability, so 
constraints on cosmological parameters can not be read directly
from these figures.  To read off constraints on individual cosmological
parameters directly from a figure, the one-dimensional figure
must be considered.  For the radio galaxy method, these are
presented in figures 4 and 5, and in 
GDW00 and DG01a.   The radio galaxy sample includes 20 FRIIb
galaxies for which $D_*$ has been computed, and the 
70 FRIIb radio galaxies in the parent population.  The 20 
sources for which $D_*$ has been determined are a subset of
the 70 sources that make up the parent population.  

The fact that the methods constrain different parts of the
$\Omega_m$-$\Omega_{\Lambda}$ plane is related to the fact
that the methods cover different redshift ranges (e.g. 
Reiss 2000).  Most of the supernovae data points are
at redshifts less than about 1 or so.  The radio galaxies
are primarily at redshifts from 0.5 to 2, and the cosmic microwave
background is at a redshift of about a thousand.  Thus, the three
data sets are complementary in their redshift coverage, and
the constraints they impose.    

\begin{figure}
\centering
\epsfig{file=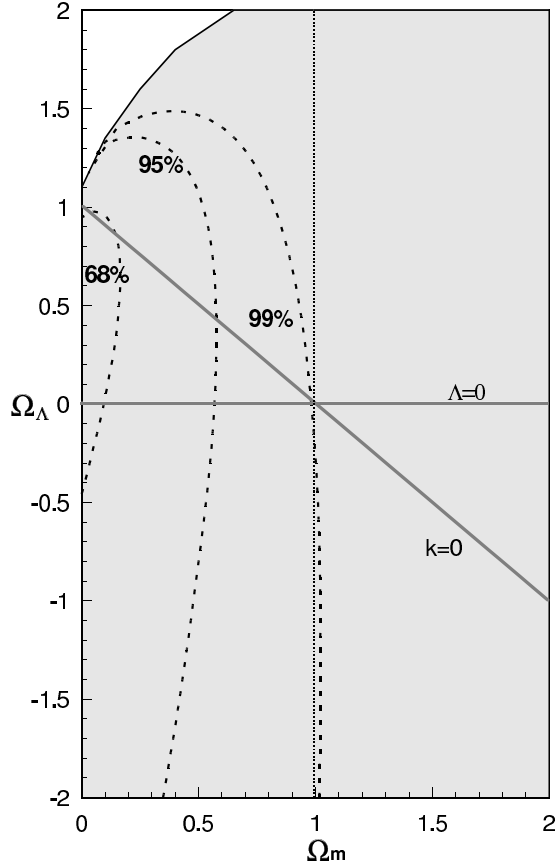,height=4.5truein}
\caption{Constraint obtained using FRIIb radio galaxies.
The projections of the
68\%, 95\%, and 99\% confidence intervals
onto either axis ($\Omega_m$ or $\Omega_{\Lambda}$)
indicates the probability associated with the
range in that one parameter, independent of
all other parameter choices.  That is, these are the
one-dimensional confidence contours (see GDW00).}
\label{1d}
\end{figure}

\begin{figure}
\centering
\epsfig{file=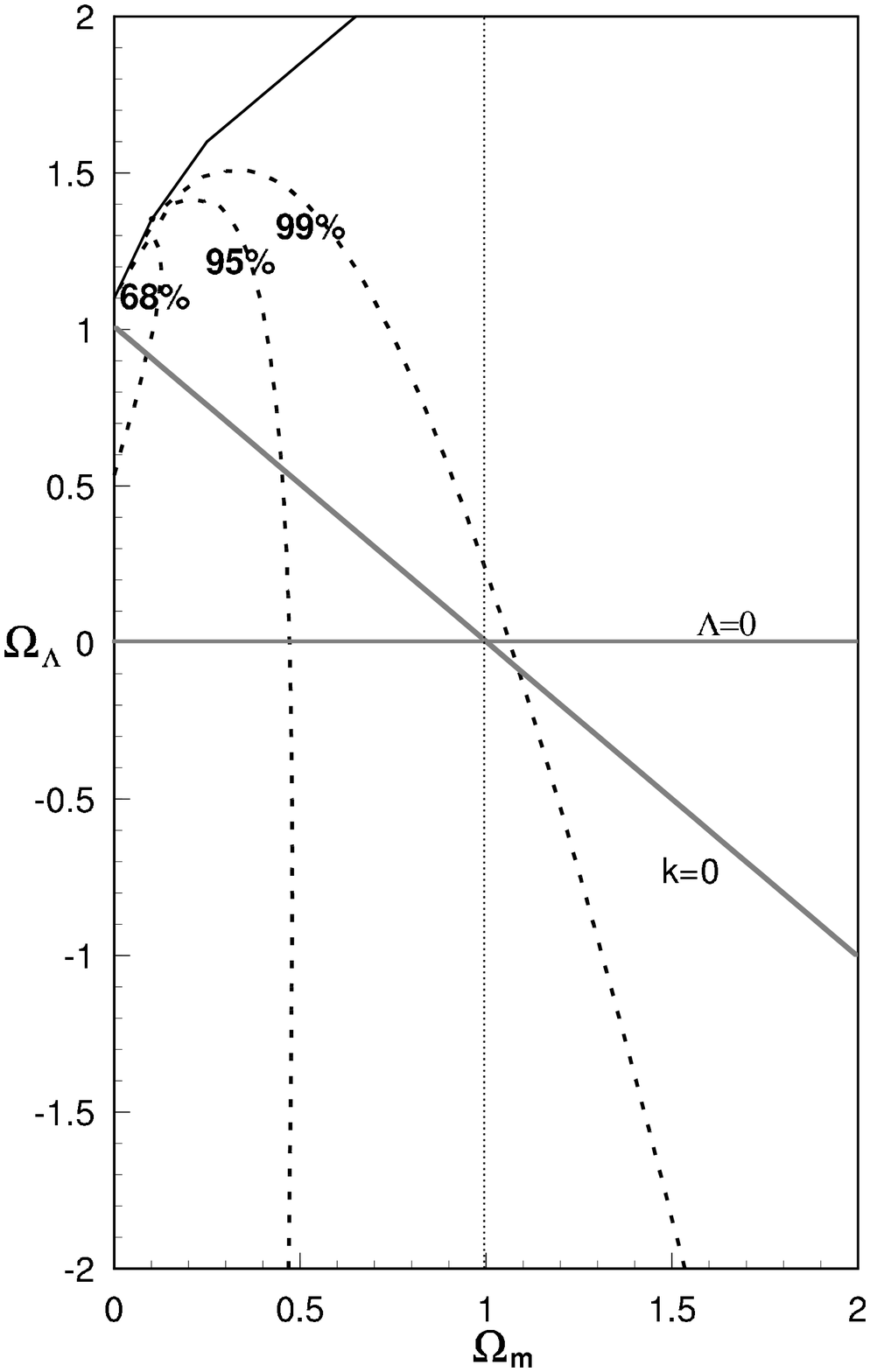,height=4.5truein}
\caption{The same as Figure \ref{1d}, excluding 
the lowest redshift bin (i.e.  
Cygnus A), from the fit.  Only sources with redshifts
greater than 0.3 are included in this analysis (see GDW00).
The results are nearly identical to those presented in figure 4.}
\label{nocyg1d}
\end{figure}

\section{FRIIb Radio Galaxies as a Modified Standard Yardstick}

A detailed description of FRIIb radio galaxies as a tool to determine
global cosmological parameters may be found in Daly (1994), 
Guerra (1997), Guerra \& Daly (1998), GDW00, DG01a,b.   

FRIIb radio galaxies have very regular bridge structure. They
are the most powerful FRII sources, having radio powers about a
factor of 10 above the classical FRI-FRII separation.  The regular
bridge structure indicates that both the instantaneous and
time-average rate of growth of the source is supersonic
(e.g. Daly 2001).  

The method is based on the following premises and assumptions:
the forward region of 
FRIIb radio galaxies are governed by strong shock physics; 
the total source lifetime $t_*$ is related to the 
beam power $L_j$ via the relation $t_* \propto L_j^{-\beta/3}$; and
the sources in the parent population at a given redshift
have a similar maximum or average size.  If these conditions
are satisfied, then FRIIb
radio galaxies provide a reliable probe of global cosmological
parameters.  Observations of FRIIb sources indicate that the
forward region (near the radio hotspot) are governed by strong
shock physics.  A power-law relation between the beam power
and the source lifetime is expected/predicted in currently
popular models to produce large scale jets (DG01b). 
And, the dispersion in source size at a given redshift
for the parent population of sources suggests that the sources
at a given redshift do have a similar average size (see figure 6).  

From these simple assumptions, the average size
a given source will have if it could be observed over its
entire lifetime is 
\begin{equation}
D_* \propto (P_La_L^2)^{-\beta/3}~
v_L^{1-\beta/3} \propto (a_or)^{-2\beta/3 +4/7} ~,
\end{equation} 
where each contributor to $D_*$ can be determined from 
radio bridge observations including the lobe pressure
$P_L$, the lobe width $a_L$, and the average rate at which
the bridge is lengthening $v_L$.  

Minumum energy conditions are not assumed to hold
in the bridges of the sources; an offset
from minimum energy conditions is included, and is
one of the parameters that is obtained from the analysis
(see Wellman, Daly, \& Wan 1997b [WDW97b]).  

\begin{figure}
\centering
\epsfig{file=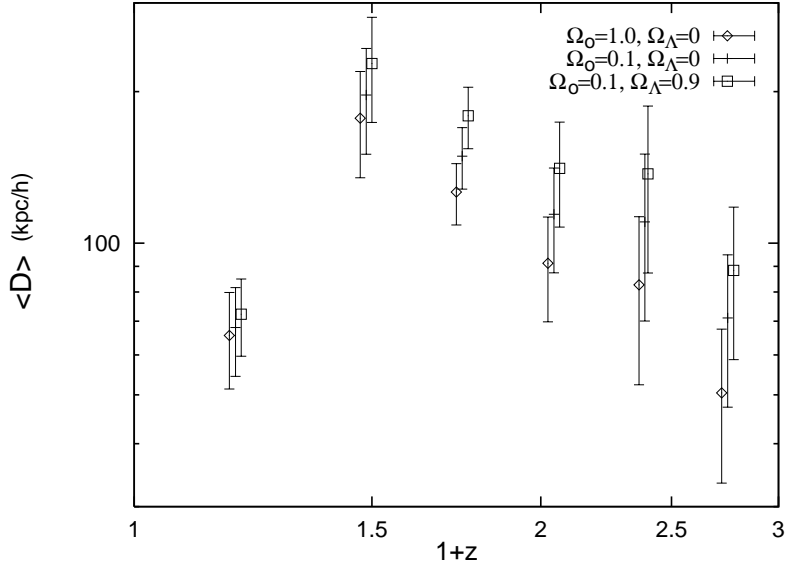,height=3truein}
\caption{The average lobe-lobe size $\langle D \rangle$ of
powerful extended 3CR radio galaxies for the 70 FRIIb radio galaxies
that comprise the parent population of radio galaxies. Different
choices of cosmological parameters are shown (here $\Omega_m = \Omega_o$).  
Note that $<D> \propto (a_or)$.}
\label{dave}
\end{figure}

\begin{figure}
\centering
\epsfig{file=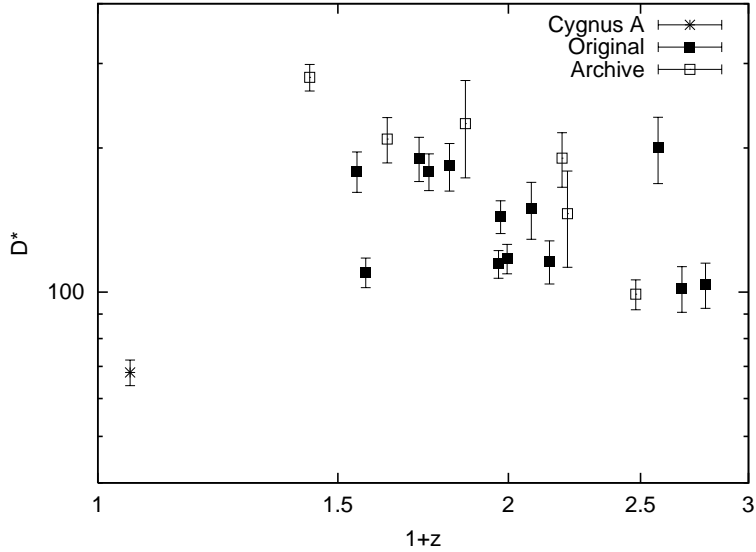,height=3truein}
\caption{The model dependent, average size $D_*$ the source
would have if it could be observed over its entire lifetime 
computed assuming $\beta=1.75$, $b=0.25$, $\Omega_m=0.1$, and
$\Omega_{\Lambda}=0$; b is the offset of the bridge magnetic field
strength from the minimum energy value. Note that $D_* \propto (a_or)^{-0.6}$.}
\label{dstar}
\end{figure}

The average size of a given source $D_*$ is assumed to be
equal to the average size of the parent population at the
same redshift $<D>$, where $<D> \propto (a_or)$.  
Thus, the ratio $<D>/D_*$, which must be a constant,
can be used to determine the coordinate distance to
the source since 
\begin{equation}
<D>/D_* \propto (a_or)^{2\beta/3+3/7}~.
\end{equation}

Figure 6 shows $<D>$ as a function of redshift, and
figure 7 shows $D_*$ as a function of redshift.  The 
method relies on the fact that $D_* \propto (a_or)^{-0.6}$ 
for $\beta = 1.75$, while $<D> \propto (a_or)$.  Thus,
the method boils down to finding the cosmological parameters
such that the shapes of the best-fitting curves to $D_*(z)$
and $<D>(z)$ match.  That is why the method is independent
of the local distance scale and of local sources.  
Differences between cosmological models begin to be obvious
at redshifts greater than about 0.5;  As shown in figure 6,
the slope of the line that describes $<D>(z)$ steepens 
as the cosmological paramaters move from being dominated
by a cosmological constant to an open universe to a 
matter-dominated $\Omega_m = 1$ universe; this arises
because $<D> \propto (a_or)$.  However, the slope
of the line that describes $D_*(z)$ begins at its
steepest and becomes less steep
as the cosmological paramaters move from being dominated
by a cosmological constant to an open universe to a 
matter-dominated $\Omega_m = 1$ universe; this arises
because $<D> \propto (a_or)^{-0.6}$.  Thus, the power of
the method lies in the fact that only for the correct choice
of cosmological parameters will the shapes of $D_*(z)$ and
$<D>(z)$ be the same.  The normalization of the fits is
allowed to float; that is $<D>/D_*$ = constant, and the
constant is an output of the fit.  

\begin{figure}
\centering
\epsfig{file=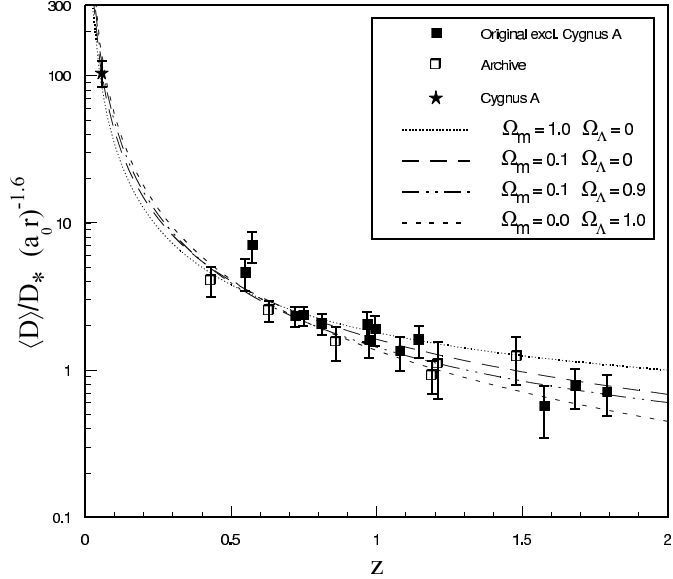,height=3truein}
\caption{The quantity $(\langle D \rangle / D_*) (a_o r)^{-1.6}$,
computed assuming $\beta=1.75$ and $b=0.25$, where b parameterizes the
offset from minimum energy conditions.  
Best fit
models for different choices of $\Omega_m$ and $\Omega_{\Lambda}$
are shown, where sources with redshifts less than 0.3 have
been {\bf excluded} from the fits.  Note that the fits are not
normalized using local or low-redshift sources, yet all of the
fits go right through the low-redshift point at a redshift
of about 0.06 (Cygnus A) and $<D>/D_*(a_or)^{-1.6} \sim 100$.
The normalization is allowed to float for each fit.}
\label{doverd}
\end{figure}

The ratio of $<D>/D_*$ is shown in figure 8, where the
ratio is multiplied by $(a_or)^{-1.6}$ to remove the 
dependence of the ratio on cosmological parameters; a 
value of $\beta=1.75$ was adopted for illustrative
purposes.  The fits shown are obtained using only radio
data with redshifts greater than 0.3, thus, this is
independent of the local distance scale and the properties
of local sources.  The best fitting lines pass right
through the data point at a redshift of about 0.06 (Cygnus A)
and $<D>/D_*(a_or)^{-1.6} = 100$, which shows that the
method is working quite well.

Evolution of source properties with redshift are much
less of a concern for the radio galaxy method than for methods that
rely upon the properties of local sources because 
the method relies on a comparison of individual
source properties with the properties of the parent population.

\section{Quintessence in a Spatially Flat Universe}

The application of FRIIb radio galaxies to constrain 
the current value of the normalized mean mass-energy density
$\Omega_Q$ and (time-independent) equation of state $w=P/\rho$ of 
quintessence 
in a spatially flat universe, $\Omega_Q=1-\Omega_m$, 
is presented by DG01a,b.  
The results are summarized here.

A universe with two primary components at the present epoch (non-relativistic
matter and quintessence) will be accelerating in its expansion
when 
\begin{equation}
1+3w(1-\Omega_m) <0~
\end{equation}
(DG01a).  

\begin{figure}
\centering
  \epsfig{file=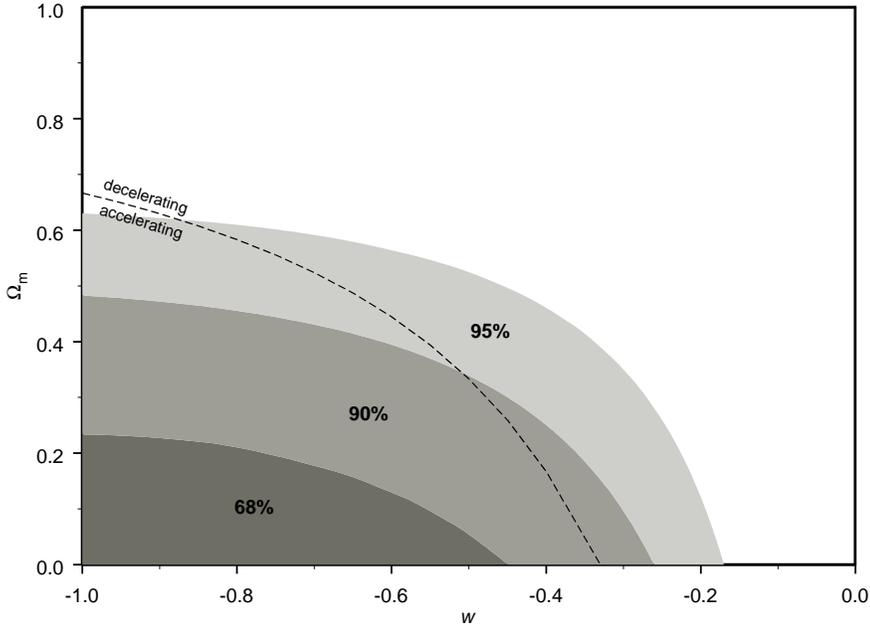,height=4truein}
  \caption{One-dimensional constraints 
on $\Omega_m$ and $w$ obtained using $D_*$ for 
20 FRIIb radio galaxies, and assuming
zero space curvature (see DG01a).}
  \label{quint1d}
\end{figure}

Figure \ref{quint1d} shows the one-dimensional constraints obtained
using a sample of 20 FRIIb radio galaxies, which are a subset
of the 70 FRIIb radio galaxies in the parent population.  Most of
the sources in the sample have redshifts between 0.5 and 2, and
identical results obtain with and without the lowest redshift bin;
again, the method is independent of the local distance scale and
of the properties of local sources.  Clearly, FRIIb radio galaxies indicate
that the universe is likely to be accelerating in its expansion
at the present time; this is about a 2 $\sigma$ result.  

In any method, it is important to consider any possible
covariance between constraints placed on different sets 
of parameters.  For the FRIIb radio galaxy method, it is
important to determine if there is any covariance between
the global cosmological parameter $\Omega_m$ and the 
model parameter $\beta$.  Figure 10 shows that there
is no covariance between these parameters.  

\begin{figure}
\centering
\epsfig{file=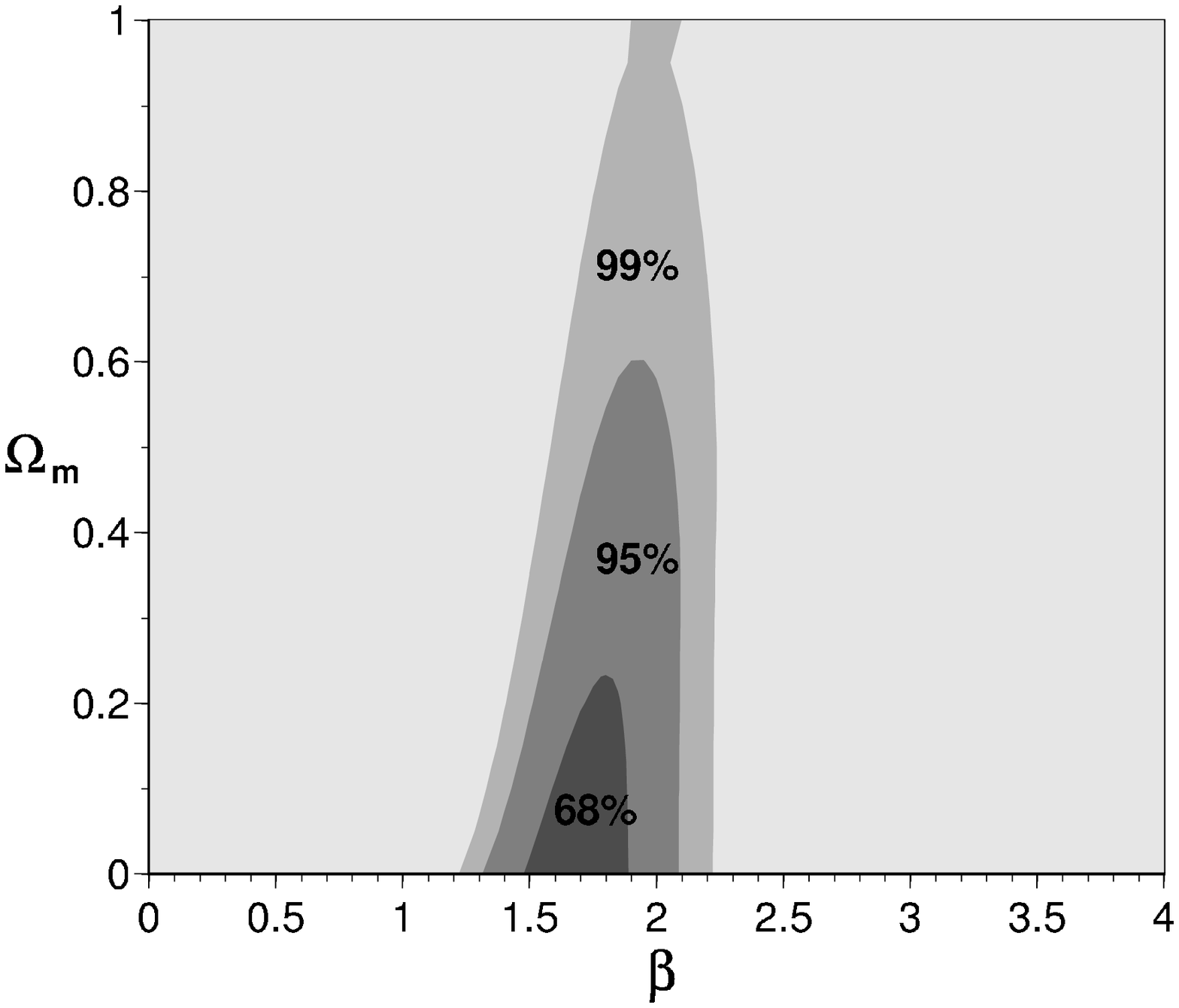,height=3truein}
\caption{One-dimensional constraints on 
$\Omega_m$ and $\beta$ for quintessence models, assuming
zero space curvture (see DG01a).}
\label{qombet}
\end{figure}

It is also important to insure that there is no covariance
between the equation of state of quintessence $w$ and
the model parameter $\beta$.  Figure 11 shows that there
is no covariance between these parameters.

\begin{figure}
\centering
\epsfig{file=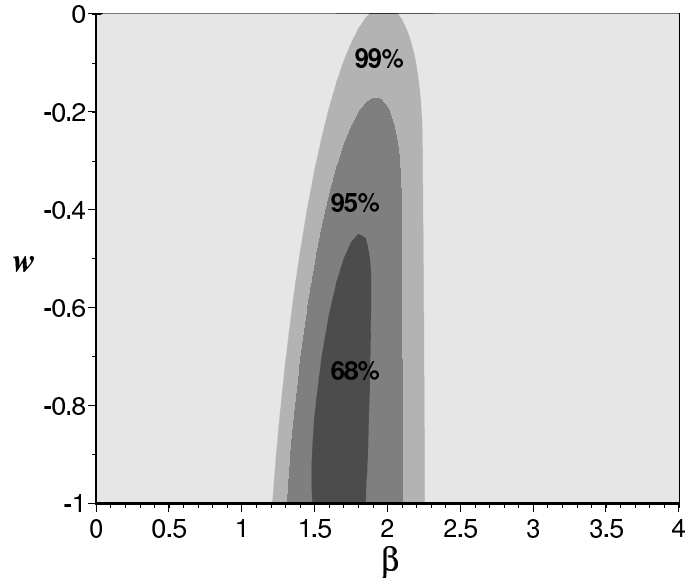,height=3truein}
\caption{One-dimensional constraints on $w$ and $\beta$ for
quintessence models, assuming
zero space curvature (see DG01a).}
\label{qwbet}
\end{figure}

Clearly, a cosmological constant, which has $w=-1$, is consistent
with the data. 

FRIIb radio galaxies can also be used to constrain global cosmological
parameters allowing for space curvature, non-relativistic matter, 
and a cosmological constant (GDW00).  
The results indicate that $\Omega_m$ must be less than one,
and $\Omega_m = 1$ is ruled out at about 99 \% confidence.  
Figures 12 and 13 show that there is no covariance between the 
model parameter $\beta$ and the cosmological parameter $\Omega_m$
allowing for a cosmological constant, or space curvature.

\begin{figure}
\centering
\epsfig{file=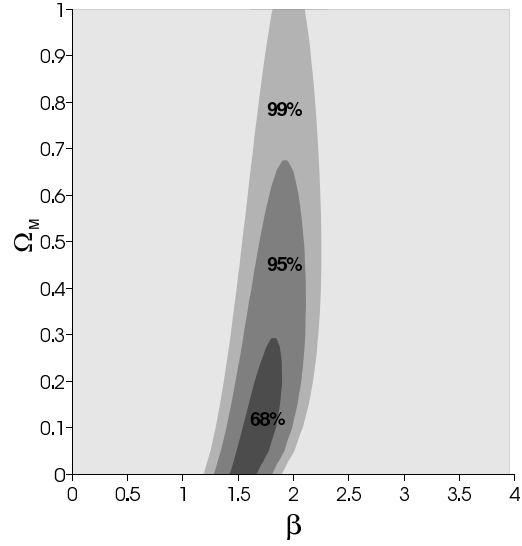,height=3truein}
\caption{One-dimensional constraints on $\Omega_m$ 
and $\beta$ assuming zero space curvature
and allowing for a cosmological constant (see GDW00).}
\label{ombetflat}
\end{figure}

\begin{figure}
\centering
\epsfig{file=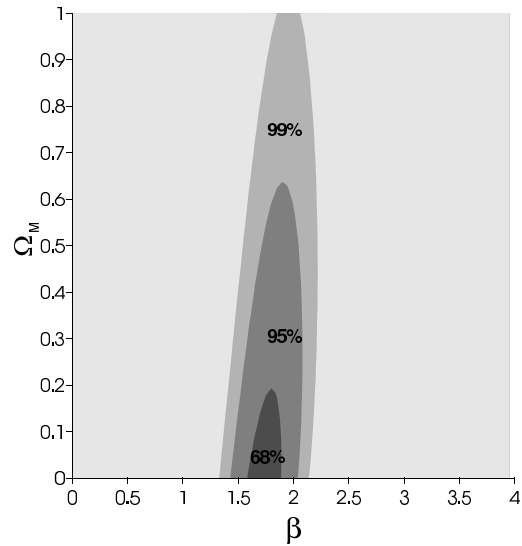,height=3truein}
\caption{One-dimensinal constraints on $\Omega_m$ and $\beta$ 
assuming no cosmological constant and allowing for space curvature
(see GDW00).}
  \label{ombetnolam}
\end{figure}

\section{Implications for the Beam Power - Total Energy Connection}

Studies of FRIIb radio galaxies indicate that the total
time $t_*$ for which the central black hole produces collimated
jets with beam power $L_j$ that ultimately power the strong shocks
near the hotspots and lobes of a given source are described
by 
\begin{equation}
t_* \propto L_j^{-\beta/3}~.  
\label{lifetime}
\end{equation}
The total energy 
$E_* = L_jt_* \propto L_j^{1-\beta/3}$, and the 
beam power is related to the total energy available
to power the outflow 
\begin{equation}
L_j \propto E_*^{3/(3-\beta)}~.
\label{beampower}
\end{equation}  
Thus, $L_j \propto E_*^{2~ to~ 3}$ for $\beta = 1.5$ to 2;
recall that $\beta = 1.75 \pm 0.25$.

An analysis of the implications for models
of jet production is given by DG01b,
and is briefly summarized here. 
If the jets are powered by the electromagnetic
extraction of the rotational energy of a spinning black
hole (Blandford 1990), then equation 
\ref{lifetime} and 
\ref{beampower}
implies that the magnetic field $B$ must be related to
the black hole mass $M$, the spin angular momentum
$S$ per unit mass $a=S/(Mc)$, and the effective size
of the black hole $m=GM/c^2$, via the relation
\begin{equation}
B \propto M^{(2\beta-3)/2(3-\beta)}(a/m)^{\beta/(3-\beta)}~,
\label{blandford}
\end{equation}
as detailed by DG01b.
Equation \ref{blandford} takes a particularly simple
form for $\beta = 1.5$; in this case, 
$B \propto (a/m)$.  For a value of $\beta$ of
1.75, equation \ref{blandford} implies that
$B \propto  M^{0.2}(a/m)^{1.4}$, and for 
$\beta = 2$, $B \propto M^{1/2}(a/m)^2$.  
Specific values for B, $L_j$, and $E_*$ are described
below, where this model is discussed further.

If the jets are powered by radiation that is
Eddington limited so $L_j = \eta_L L_E \propto \eta_L M$,
where $L_E$ is the Eddington luminosity, 
$\eta_L$ is the efficiency with which the 
radiant luminosity is converted into beam power, 
$M$ is the black hole mass, and the 
accreted mass $M_{acc}$ 
is converted to energy with an
efficiency $\eta_E$ so that $E_* = \eta_E M_{acc} c^2$
(e.g., Krolik 1999), then equation \ref{beampower}
implies that 
\begin{equation}
\eta_L \propto \eta_E^{3/(3-\beta)}(M/M_{acc})^{-3/(3-\beta)} 
M^{\beta/(3-\beta)}~.
\end{equation}
For $\beta =1.5$, this reduces to $\eta_L \propto \eta_E^2
(M/M_{acc})^{-2}M$.    
The beam power could be as high as the Eddington luminosity,
or could be a constant fraction of the Eddington luminosity,
in which case
\begin{equation}
\eta_E \propto (M/M_{acc})M^{-\beta/3}~,
\end{equation}
or $\eta_E \propto M^{-1/2}$ for $\beta=1.5$ and
$M_{acc} \sim M$.

\begin{figure}
\centering
\begin{turn}{-90}
\epsfig{file=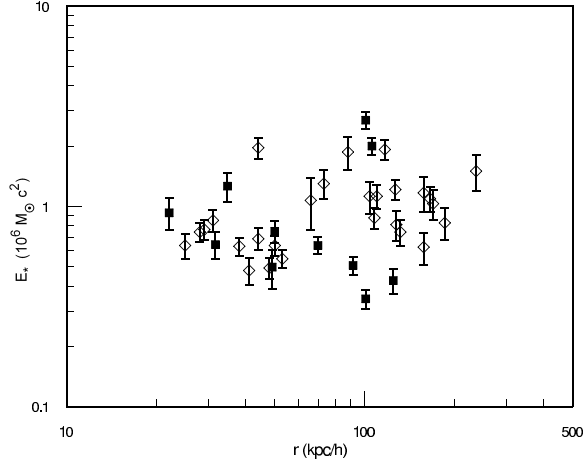,height=3truein}
\end{turn}
\caption{The total energy processed through each beam
of each source as a function of core-hotspot separation; open
symbols represent the 14 radio galaxies in the original sample
(see GD98), and
the filled symbols represent the 6 radio galaxies obtained
from the VLA archive (see GDW00).  Most sources have 2 data points, one from 
each side of the source.}
\label{esr}
\end{figure}

\begin{figure}
\centering
\begin{turn}{-90}
\epsfig{file=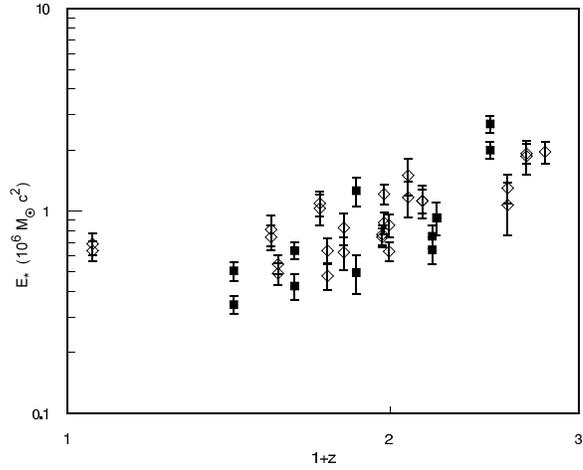,height=3truein}
\end{turn}
\caption{The total energy processed through each jet of 
each source as a function of redshift.  The symbols are the same
as in figure \ref{esr}.}
\label{esz}
\end{figure}

The total energy that will be processed by 
each source through each of its two large-scale jets over its lifetime
$E_*$ is shown as a function of core-hotspot separation
in Figure \ref{esr} and as a function of redshift in Figure \ref{esz}.  
 
\begin{figure}
\centering
\begin{turn}{-90}
\epsfig{file=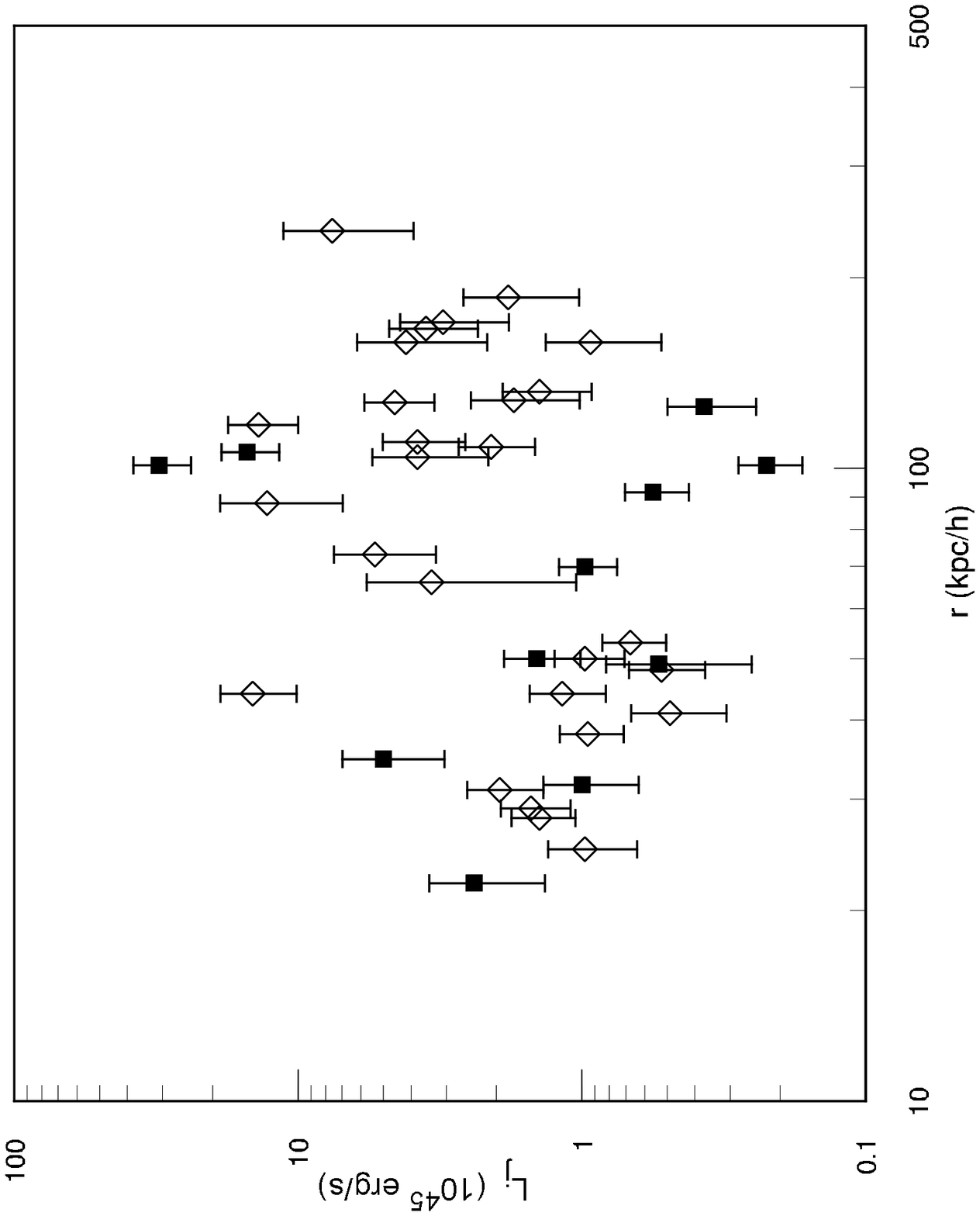,height=3truein}
\end{turn}
\caption{The beam power processed through 
each side of each source as a function of core-hotspot separation.
The symbols are the same
as in figure \ref{esr}.}
\label{ljr}
\end{figure}

\begin{figure}
\centering
\begin{turn}{-90}
\epsfig{file=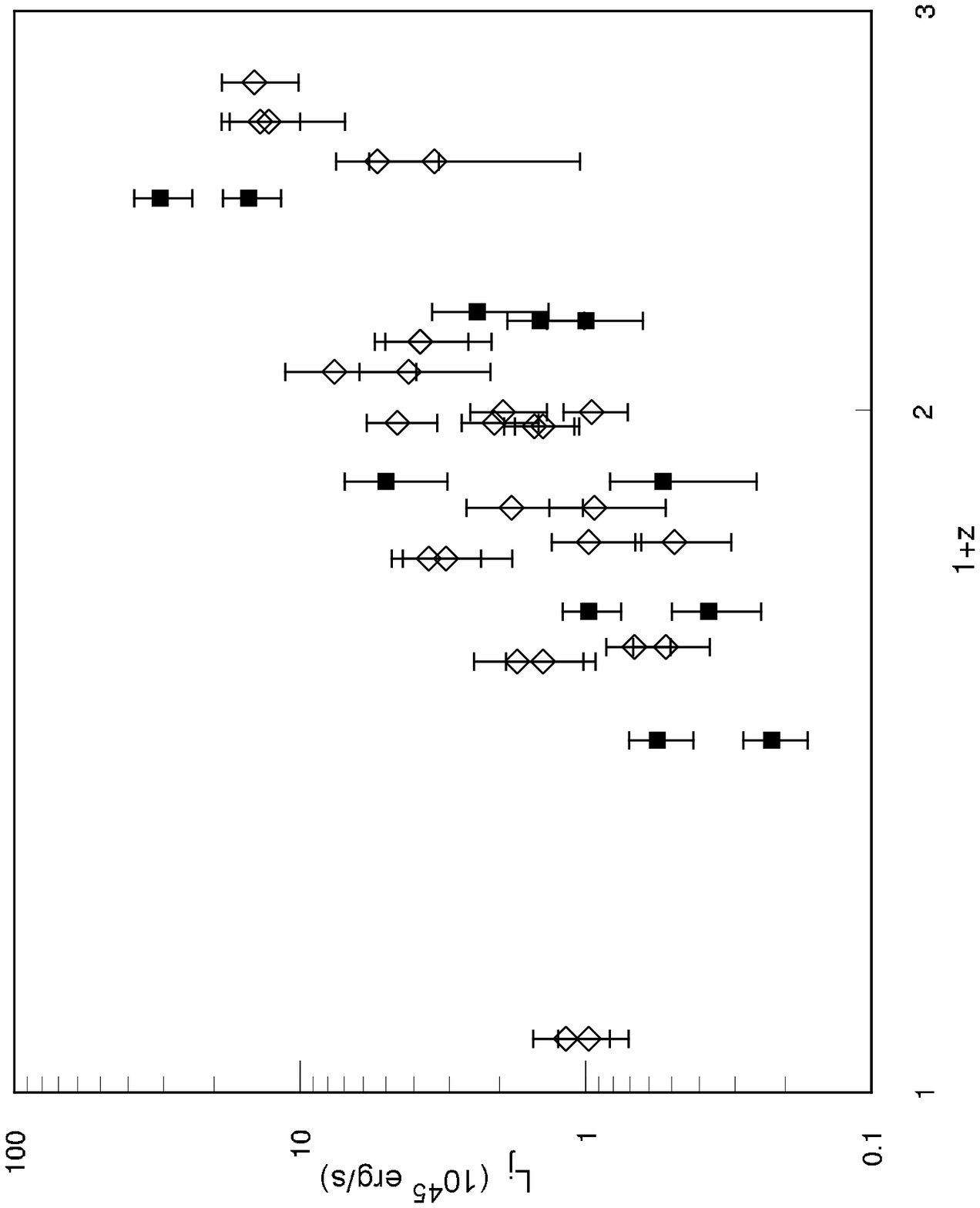,height=3truein}
\end{turn}
\caption{The beam power processed through each side of
each source as a function of redshift.
The symbols are the same
as in figure \ref{esr}.}
\label{ljz}
\end{figure}

The beam powers of the sources are shown in Figure
\ref{ljr} as a function of core-hotspot separation, and
in Figure \ref{ljz} as a function of redshift. 

Clearly, the beam power increases with redshift.
However, if the beam power is related to the
Eddington luminosity, then it is proportional to the
black hole mass, which is expected to increase
with time, or decrease with redshift.  
In addition, the beam powers
are typically about $10^{45}$ erg/s, implying a black
hole mass of only about $10^7 M_{\odot}$, which is 
smaller than the black hole masses associated with 
galaxies in the cores of clusters of galaxies.  
The scaling of the beam power, total energy, and
total source lifetime as given above, and can
be normalized using the fact that when the beam
power is 
$L_j \sim 10^{45} erg/s$,
the total active lifetime is $t_* \sim 10^7$ yr, and
the total energy processed through the jet is
$\sim 5 \times 10^5 c^2 M_{\odot}$ (see DG01b, or
figures 14, 16, and 18).  

Since the sources seem to be preferentially located
near the cores of clusters or protoclusters of
galaxies (see Wellman, Daly, \& Wan 1997a [WDW97a] and 
WDW97b), a model in which a massive
black hole acquires substantial rotational energy
after coallescing with another massive black
hole, such as that of Wilson \& Colbert (1995) is quite attractive.
A timescale can be constructed by dividing
$E_*$ by $L_{EM}$; using equations (3.38) and
(3.39) from Blandford (1990), the timescale of
the outflow is of order $10^9/(M_8 B_4^2)$ yr,
where $M_8$ is the mass of the black hole in
units of $10^8 M_{\odot}$, and $B_4$ is the magnetic
field strength in units of $10^4$ G.  A total lifetime
of about $10^7$ yr results for a black hole mass
of about $10^9 M_{\odot}$ and a field strength
of about $3 \times 10^4$ G.  These values would
lead to beam powers of about $10^{45}$ erg/s and
total energies of about $10^{60}$ erg, equivalent
to a rest mass of about $5 \times 10^5 M_{\odot}$,
for $(a/m) \sim 1/30$, which are reasonable
numbers, and which agree with the results obtained
empirically.  

For comparison, the total lifetimes
of the 20 sources studied here are shown in figure 18
as a function of core-hotspot separation, and in figure 19
as a function of redshift.  Clearly, high-redshift
sources have shorter total lifetimes.

\begin{figure}
\centering
\begin{turn}{-90}
\epsfig{file=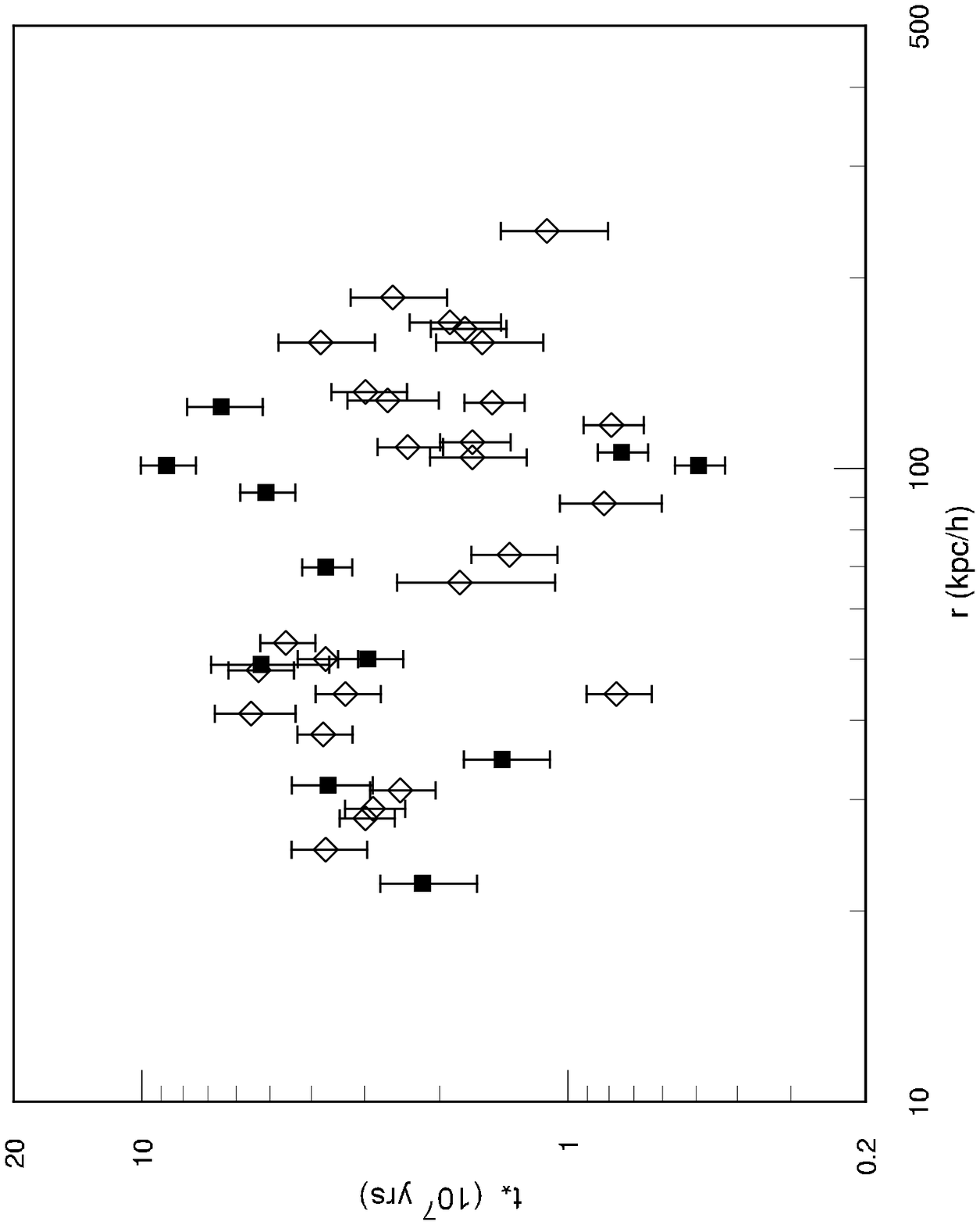,height=3truein}
\end{turn}
\caption{The total jet lifetime for each side of 
each source as a function of core-hotspot separation.
The symbols are the same
as in figure \ref{esr}.}
\label{tjr}
\end{figure}

\begin{figure}
\centering
\begin{turn}{-90}
\epsfig{file=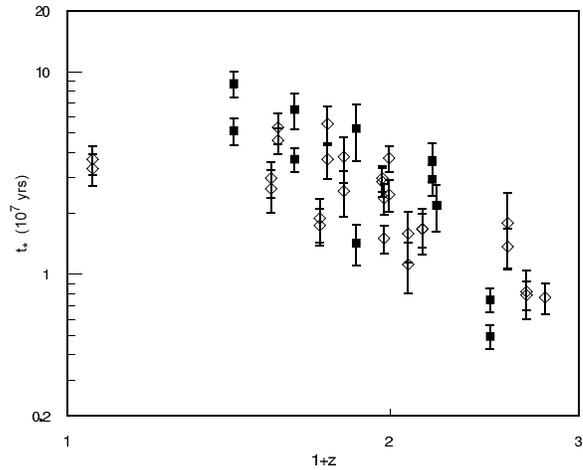,height=3truein}
\end{turn}
\caption{The total jet lifetime for each side of each source 
as a function of redshift.  The symbols are the same
as in figure \ref{esr}.}
\label{tjz}
\end{figure}

\section{FRIIb Radio Sources Provide Probes of Evolution of Structure}

The structure of the radio bridges of FRIIb sources can be used to
determine the ambient gas pressure, density, temperature, 
and the overall Mach number of the forward region 
of the source, as detailed in a series of papers (Wan \& Daly 
1998; WDW97a,b; 
WDG00).  The use of the properties of the radio
bridge to determine the ambient gas pressure does not rely
upon the assumption of minimum energy conditions, and it is
independent of the rate of growth, or lobe propagation velocity,
of the source, and hence is independent of any aging analysis applied
to the radio bridge.  The sources are in gaseous environments with
pressures and composite pressure profiles like those found in 
clusters of galaxies, though some redshift evolution of the 
ambient gas pressure is suggested by the results obtained. 

The ambient gas density may be obtained by studying
the ram pressure confinement of the forward region of
the radio source.  This is done by combining the
rate of growth of the source, determined using a spectal aging
analysis across the radio bridge of the source, with the
pressure near the end of the source, as done by 
Perley \& Taylor (1990) for 3C295, Carilli et al. (1990) for
Cygnus A, WDW97a 
for a sample of 14 radio galaxies and
8 radio loud quasars, and GDW00 for 
a sample of 6 radio galaxies.
These studies indicate that the sources are in cluster-like
gaseous environments, though some redshift evolution of the
density and density profile is indicated by the study of 
WDW97b.  

The shape of the radio bridge leads to a determination
of the average Mach number with which the forward region
of the source moves into the ambient gas (WDW97a).
The Mach number can be combined with the rate of
growth of the source, or lobe propagation velocity, to
obtain the temperature of the ambient gas (WDW97a).
This analysis also indicates that the sources are in 
gaseous environments with temperatures similar to those in present
day cluster of galaxies.

Thus, there are some indications from the properties of 
FRIIb radio sources that they are in the cores of clusters
or proto-clusters of galaxies, and 
that the properties of the gas in the cores of galaxy clusters
evolves from a redshift of zero to a redshift of two.

\section{Conclusions}

FRIIb radio galaxies provide a probe of the 
coordinate distance to very high redshift sources, including
sources with redshifts between one and two.  The results
obtained using FRIIb radio galaxies are consistent with
those indicated by measurements of the cosmic microwave
background (de Bernardis et al.\ 2000, Balbi et al.\ 2000, Bond et al.\ 2000), 
and type Ia supernovae (Riess et al.\ 1998, Perlmutter et al.\ 1999).
The modified standard yardstick method, which uses FRIIb
radio galaxies, is complementary to the modified standard
candle method, which uses type Ia supernovae.  Possible sources
of error in each method are likely to be very different.
Thus, the fact that they yield consistent results suggests that
any errors not yet accounted for in either method must be small compared with 
errors that are currently known and accounted for.

FRIIb radio galaxies indicate that $\Omega_m$ must be low;
$\Omega_m = 1$ is ruled out at about 99 \% confidence.
Measurements of the cosmic microwave background radiation indicate
that the universe has zero space curvature (e.g., Bond et al.\ 2000).  In 
a universe with zero space curvature and quintessence with
equation of state $w$, the modified standard yardstick 
method using FRIIb radio galaxies indicates that it is likely
that the expansion of the universe is accelerating at present.
The expansion rate of the universe will be 
accelerating when
\begin{equation}
1+3w(1-\Omega_m) <0~,
\end{equation} 
DG01a.  This line is 
drawn on Figure 9; the region below the line indicates 
that the universe is accelerating at present; those above
the line indicate a decelerating universe.  Radio galaxies
alone indicate that the universe is accelerating in its
expansion at the current epoch; this is an 84 \% confidence 
result.   

The application of the modified standard yardstick method not
only allows a determination of global cosmological parameters,
it also allows a determination of the model parameter $\beta$; current
results indicate that $\beta$ is about $1.75 \pm 0.25$.  The parameter
$\beta$ can be used to constrain models of energy extraction
from the central massive object, presumed to be a massive
black hole; these constraints are 
described by DG01b and are summarized
in section 5.  The value of $\beta$ determined empirically  
is expected/predicted in models where jet formation, power, 
and energy are related to the electromagnetic extraction of
the rotational energy of a spinning black hole, and is consistent
with models in which jet production is related to the 
Eddington luminosity of the black hole region.  

Independent of the application of FRIIb radio galaxies to constrain
cosmological parameters and models of energy extraction from massive
black holes, FRIIb radio galaxies can also be used to study the 
gaseous environments of this type of AGN.  The sources may be
used to study the pressure, density, and temperature of the gas
around them, and appear to be located in the cores of clusters
or protoclusters of galaxies, as summarized in section 6.  

FRIIb radio sources appear to be governed by strong shock
physics, and hence are in a regime that is easy to quantify.
This fact, coupled with the interesting relation between
the beam power and total energy that all of the sources seem
to follow, makes them ideally suited to cosmological studies.
In addition to their use as a modified standard yardstick, they
also provide a probe of models of evolution of structure through
their use to study the properties of the cores of clusters or
proto-clusters of galaxies. The fact that they are located near the
centers of clusters or proto-clusters of galaxies is probably
related to the similar physical mechanism of jet formation and
energy extraction.

\acknowledgments

It is a pleasure to thank Megan Donahue, Paddy Leahy, Chris O'Dea, Adam Reiss,
and Max Tegmark for helpful comments and discussions.  We are
grateful to Lin Wan and Greg Wellman for their contributions
to the study of FRIIb radio souces.  
This research was supported in part by National Young 
Investigator Award AST-0096077 
from the US National Science Foundation, and
by the Berks-Lehigh Valley College of Penn State 
University.   
Research at Rowan University was supported in part by the College
of Liberal Arts and Sciences and National Science Foundation
grant AST-9905652.

\end{document}